\documentclass[a4paper,11pt]{article}

\usepackage{amsmath}
\usepackage{amssymb}
\usepackage{amscd}
\usepackage{latexsym}
\usepackage{epsfig}
\usepackage{bbm}

\newtheorem{defi}{Definition}

\newtheorem{satz}[defi]{Theorem}

\newcommand{\tr}{{\operatorname{Tr}\,}}

\newcommand{\bra}[1]{{\langle{#1}|}}
\newcommand{\ket}[1]{{|{#1}\rangle}}
\newcommand{\ketbra}[1]{{\ket{#1}\bra{#1}}}
\newcommand{\1}{{\mathbbm{1}}}
\newcommand{\C}{{\mathbb{C}}}

\newcommand{\alg}[1]{{\mathfrak{#1}}}
\newcommand{\fset}[1]{{\mathcal{#1}}}

\newlength{\blank}
\begin{document}

\title{On the fidelity of two pure states}
\author{Andreas Winter\thanks{Address: SFB 343, Fakult\"at f\"ur Mathematik, Universit\"at Bielefeld, Postfach 100131, D--33501 Bielefeld, Germany. Email: \texttt{winter@mathematik.uni-bielefeld.de}. Research supported by SFB 343 ``Diskrete Strukturen in der Mathematik'' of the Deutsche Forschungsgemeinschaft.}}
\date{November 13, 2000}

\maketitle


\begin{abstract}
  The \emph{fidelity} of two pure states
  (also known as \emph{transition probability}) is a symmetric
  function of two operators, and well--founded operationally
  as an event probability in a certain preparation--test pair.
  Motivated by the idea that the fidelity is the continuous
  quantum extension of the combinatorial equality function,
  we enquire whether there exists a \emph{symmetric operational}
  way of obtaining the fidelity. It is shown that this is
  impossible.
  Finally, we discuss the optimal universal approximation
  by a quantum operation.
\end{abstract}


\section{Introduction}
\label{sec:intro}
For two pure quantum states $\pi=\ket{\varphi}\bra{\varphi}$ and
$\tau=\ket{\theta}\bra{\theta}$ on the space ${\cal H}$,
which we assume throughout to be of dimension $d<\infty$,
the (pure state) fidelity is
$$F(\pi,\tau)=\tr \pi\tau=|\langle\varphi\ket{\theta}|^2.$$
Its operational justification is as follows: suppose we
test the system for being in state $\tau$, described by
the projection valued measure (PVM) $(\tau,\1-\tau)$,
then the probability of an
affirmative answer, the actual preparation being $\pi$,
is $F(\pi,\tau)$. It is one of the features of quantum theory
that the same probability arises if the system is prepared in
state $\tau$, and is tested for $\pi$, see the discussion
in~\cite{peres:concepts}, chapter 2. This is reflected in the
symmetry of $F$: $F(\pi,\tau)=F(\tau,\pi)$.
\par
By restricting attention to a set of orthonormal vectors
$\ket{x}$, $x\in\fset{X}$, one has
\begin{equation*}
  F(\ketbra{x},\ketbra{y})=\delta_{xy}
                          =\begin{cases}
                             1 & \text{ if }x=y \\
                             0 & \text{ if }x\neq y.
                           \end{cases}
\end{equation*}
Thus, on $\fset{X}\times\fset{X}$, $F$ represents the test for
equality of two given elements from $\fset{X}$. Observe that this
characterization is symmetric in the two variables: we can
imagine a classical computing machine taking as input $x$ and $y$
from $\fset{X}$, which outputs $\delta_{xy}\in\{0,1\}$.

\section{The problem}
\label{sec:problem}
The question arises whether or not an operational justification
for $F$ is possible that is symmetrical, too.
Note that in the above discussion
one of $\pi,\tau$ figures as a state, whereas the other as a
projection of a test. Hence, two possibilities seem natural:
either both have to be given as quantum states, or both as
tests. In either case we want to find a procedure to sample
the binary distribution $(\tr\pi\tau,1-\tr\pi\tau)$ once, i.e.
produce the first outcome with probability $\tr\pi\tau$, and
the second with probability $1-\tr\pi\tau$.

\subsection{Two states}
\label{subsec:2states}
A would--be fidelity estimator for two unknown states is a map
$$F:\pi\otimes\tau\mapsto (\tr\pi\tau)z_1+(1-\tr\pi\tau)z_0,$$
where $z_0,z_1$ are the (orthogonal) idempotent generators of a two--dimensional
commutative algebra.\footnote{Note that the restriction to one copy
  of $\pi,\tau$, each, is crucial: if we were allowed to use the
  preparation device for $\pi,\tau$ indefinitely often,
  we can do a tomography~\cite{tomography} of the states,
  and actually \emph{compute} $\tr\pi\tau$.}
As is immediate, this is indeed uniquely extendible to
a trace preserving linear map on $\alg{L}({\cal H})\otimes\alg{L}({\cal H})$.
It is even positive --- on the separable states! But not on the whole state space:
for example consider a pure state vector
$$\ket{\psi}=\alpha\ket{e_0 f_0}+\beta\ket{e_1 f_1}$$
in ${\cal H}\otimes{\cal H}$, with unit vectors $e_0\perp e_1$, $f_0\perp f_1$,
and (w.l.o.g.) $\alpha,\beta>0$ such that $\alpha^2+\beta^2=1$.
Then
\begin{align*}
  \ketbra{\psi}&=\alpha^2    \ketbra{e_0}\otimes\ketbra{f_0}
                  +\beta^2     \ketbra{e_1}\otimes\ketbra{f_1}               \\
           &\phantom{=}+\alpha\beta \ket{e_0}\bra{e_1}\otimes\ket{f_0}\bra{f_1}
                  +\alpha\beta \ket{e_1}\bra{e_0}\otimes\ket{f_1}\bra{f_0}.
\end{align*}
Note that
\begin{align*}
  |\langle e_0\ket{f_0}|^2 &=p=|\langle e_1\ket{f_1}|^2, \\
  |\langle e_0\ket{f_1}|^2 &=q=1-p=|\langle e_1\ket{f_0}|^2.
\end{align*}
Finally, introducing
$$\langle e_0\ket{f_1}=e^{i\gamma}\sqrt{q},\quad
  \langle e_1\ket{f_0}=e^{i\delta}\sqrt{q},$$
we can calculate the $1$--component of $F(\ketbra{\psi})$:
\begin{align*}
  F(\ketbra{\psi})_{1}&= \alpha^2 p+\beta^2 p
                          +\alpha\beta\langle e_1\ket{f_0}\langle f_1\ket{e_0}
                          +\alpha\beta\langle e_0\ket{f_1}\langle f_0\ket{e_1} \\
                      &= p+2q\alpha\beta\cos(\gamma-\delta),
\end{align*}
which may obviously be negative, e.g. $F(\ketbra{\psi})_{1}=-1$ for
$p=0$, $q=1$, $\alpha=\beta=1/\sqrt{2}$, and $\gamma-\delta=\pi$.
\par
It is interesting to note that we encountered here what is called
a \emph{entanglement witness} (as introduced by
Terhal~\cite{terhal:maps}): a linear map positive on products,
but negative on certain entangled states which it ``certifies''.
The operator $W=F^*(z_1)$ (using the dual map $F^*$ of $F$ with respect to the
Hilbert--Schmidt trace pairing) is the operator
version of this entanglement witness: it has the property
$$\tr(\pi\otimes\tau W)=\tr \pi\tau\geq 0,$$
but for some entangled states it has negative expected value. One
can write out $W$ explicitely:
$$W=\sum_{s} X_s^*\otimes X_s$$
for any orthonormal basis $X_1,\ldots,X_{d^2}$ of $\alg{L}({\cal H})$.

\subsection{Two tests}
\label{subsec:2tests}
Suppose we are given the PVM $M=(\pi,\1-\pi)\otimes(\tau,\1-\tau)$ on
$\alg{L}({\cal H})\otimes\alg{L}({\cal H})$ as a block box.
What we can do is feed it with an arbitrarily prepared state,
and combine the outcomes into two groups. Observe that if we allow
multiple uses of the black box we can do a tomography of
$M$~\cite{POVM:tomography}
(dual to the tomography of states~\cite{tomography}).
This motivates the restriction to
a single application of $M$~\cite{2:times:qubit}.
\par
Preparing a state $\rho$ on $\alg{L}({\cal H})\otimes\alg{L}({\cal H})$
and using it with $M$, we are supplied with one of four
outcomes ($11,10,01,00$), after which we employ a statistical
decision rule: if $ij$ was measured, we vote for $1$ with
probability $p_{ij}\in[0,1]$. This is the most general
form of the procedure, and we can calculate
\begin{align*}
  \Pr\{1\}&=  p_{11}\tr(\rho(\pi\otimes\tau))+p_{10}\tr(\rho(\pi\otimes(\1-\tau))) \\
          &\phantom{=}+p_{01}\tr(\rho((\1-\pi)\otimes\tau))
                      +p_{00}\tr(\rho((\1-\pi)\otimes(\1-\tau)))                   \\
            &=  (p_{11}-p_{10}-p_{01}+p_{00})\tr(\rho(\pi\otimes\tau))             \\
            &\phantom{=}+(p_{10}-p_{00})\tr(\rho(\pi\otimes\1))
                        +(p_{01}-p_{00})\tr(\rho(\1\otimes\tau))
                        +p_{00} \tr\rho                                            \\
            &=  (p_{11}-p_{10}-p_{01}+p_{00})\tr(\rho(\pi\otimes\tau))             \\
            &\phantom{=}+(p_{10}-p_{00})\tr(\rho_1\pi)
                        +(p_{01}-p_{00})\tr(\rho_2\tau)+p_{00}.
\end{align*}
This is a polynomial in $\pi$ and $\tau$ with a bilinear,
a linear, and a constant part. Hence, for this to be equal
to $\tr\pi\tau$, necessarily
$$p_{10}=p_{01}=p_{00}=0,$$
forcing $p_{11}=1$ (choose $\pi=\tau$). So, we have to look for a
state $\rho$ satisfying
$$\tr\pi\tau=\tr(\rho(\pi\otimes\tau)).$$
However, by subsection~\ref{subsec:2states} there does not even
exist a solution $0\leq\rho\leq\1$ to this equation.
\par\bigskip
Our result can be understood as another new feature of quantum
information as compared to classical information: whereas there
is an identity test for classical data, symmetrical in the two
inputs, the corresponding natural quantum version, namely
the fidelity, is forbidden by the quantum mechanical laws:
not only are we unable to access the precise value of it,
we cannot even once sample the corresponding Bernoulli
variable.
\par
In fact, the proof of the following section~\ref{sec:universal} shows that
there is \emph{no} operational quantum extension of the
classical identity test at all:
\begin{satz}
  \label{satz:no-fidelity}
  There is no test $T$ on ${\cal H}\otimes{\cal H}$
  (i.e.~$0\leq T\leq\1\otimes\1$), such that for all
  states $\pi,\tau$ on ${\cal H}$
  \begin{align*}
    \tau=\pi     &\Longrightarrow \tr(\pi\otimes\tau T)=1,\\
    \tau\perp\pi &\Longrightarrow \tr(\pi\otimes\tau T)=0.\\
  \end{align*}
\end{satz}
\par
Thus, we have exhibited a new no--go theorem regarding
quantum mechanics, in the line of the no--cloning 
theorem~\cite{w:z}.

\section{Universal approximation for two states}
\label{sec:universal}
After failing to find allowed procedures to sample
the fidelity distribution $F(\pi,\tau)$, we resort to
approximate this ideal behaviour in an optimal way.
\par
To find the optimal approximation to the fidelity estimator, we have to
minimize the expression
$$\delta(A)=\max_{\pi,\tau} \left|\tr\left((\pi\otimes\tau)A\right)-\tr\pi\tau\right|$$
with respect to $0\leq A\leq\1$. We may assume that the optimal $A$ is invariant
under the actions
$$\pi\otimes\tau\longmapsto\tau\otimes\pi$$
and
$$\pi\otimes\tau\longmapsto U\pi U^*\otimes U\tau U^*,\quad U\in{\cal U}({\cal H}).$$
The reasoning is the same as for universal cloning~\cite{clone} and
Bloch vector flipping~\cite{qNOT} machines:
because of invariance of the fidelity function
and triangle inequality, an optimal solution cannot become worse
if we average it over the group action using Haar measure.
\par
Since the squared representation of the unitary group has exactly 2 irreducible
components, the \emph{symmetric} and the \emph{antisymmetric} subspace,
${\cal S}$ and ${\cal A}$, respectively, with corresponding projectors
$\Pi_{\cal S}$ and $\Pi_{\cal A}$, the most general $A$ to consider has the
form
$$A=\sigma\Pi_{\cal S}+\alpha\Pi_{\cal A},\quad 0\leq\sigma,\alpha\leq 1.$$
To evaluate $\delta(A)$ choose an orthonormal basis $e_1,\ldots,e_d$
of ${\cal H}$. Then
$${\cal S}={\rm span}\left\{f_i=e_i\otimes e_i,
               f_{ij}=\frac{e_i\otimes e_j+e_j\otimes e_i}{\sqrt{2}}: i<j\right\},$$
and note that the $f_i,f_{ij}$ form an orthonormal basis of ${\cal S}$.
\par
Now by unitary invariance we may assume that
\begin{align*}
  \pi  &=\ketbra{e_1}\text{ and} \\
  \tau &=\left(u\ket{e_1}\! +\! v\ket{e_2}\right)\left(u\bra{e_1}\! +\! v\bra{e_2}\right),
          \ u,v\geq 0,\ u^2+v^2=1.
\end{align*}
Hence, noting $\tr\pi\tau=u^2$,
\begin{align*}
  \delta(A) &=\max_{u,v} \left|\sigma\tr\left((\pi\otimes\tau)\Pi_{\cal S}\right)
                        +\alpha\tr\left((\pi\otimes\tau)\Pi_{\cal A}\right)-u^2\right|  \\
            &=\max_{u,v} \left|\alpha
                    +(\sigma-\alpha)\tr\left((\pi\otimes\tau)\Pi_{\cal S}\right)-u^2\right|,
\end{align*}
and calculating
\begin{align*}
  \tr\left((\pi\otimes\tau)\Pi_{\cal S}\right)
        &= \left\| \Pi_{\cal S}\ket{e_1}\otimes(u\ket{e_1}+v\ket{e_2}) \right\|_2^2 \\
        &= \left| \left(\bra{e_1}\otimes\bra{e_1}\right)
                  \left(\ket{e_1}\otimes(u\ket{e_1}\! +\! v\ket{e_2})\right)\right|^2  \\
        &\phantom{=}+\left| \left(\bra{e_2}\otimes\bra{e_2}\right)
                  \left(\ket{e_1}\otimes(u\ket{e_1}\! +\! v\ket{e_2})\right)\right|^2  \\
        &\phantom{=}+\left|
           \frac{\bra{e_1}\otimes\bra{e_2}\! +\!\bra{e_2}\otimes\bra{e_1}}{\sqrt{2}}
                   \left(\ket{e_1}\otimes(u\ket{e_1}\! +\!v\ket{e_2})\right)\right|^2  \\
        &= u^2+0+\frac{v^2}{2}=\frac{1+u^2}{2}
\end{align*}
we end up with
\begin{align*}
  \delta(A) &=\max_{0\leq u^2\leq 1}
               \left|\alpha+(\sigma-\alpha)\frac{1+u^2}{2}-u^2\right| \\
            &=\max_{0\leq x\leq 1}
               \left|\frac{\sigma+\alpha}{2}+\left(\frac{\sigma-\alpha}{2}-1\right)x\right| \\
            &=\max \left\{ \frac{\sigma+\alpha}{2},1-\sigma \right\}.
\end{align*}
To minimize this we have to choose $\alpha=0$ and $\sigma=2/3$.
The optimal test is thus
$$A=\frac{2}{3}\Pi_{\cal S},$$
achieving $\delta(A)=\delta_{\min}=1/3$.
\par
The general case of $n$ copies of the two states, and $m$ samples
to be produced, is discussed in the appendix.
\par\bigskip
Complementing theorem~\ref{satz:no-fidelity} above, note that we
\emph{can} obtain \emph{partial} information on the fidelity.
For example, the optimal test $T=2/3\Pi_{\cal S}$ has the property
that
\begin{align*}
  \tr(\pi\otimes\tau T)>\frac{1}{2}&\text{ iff }\ \tr\pi\tau>\frac{1}{2},\\
  \tr(\pi\otimes\tau T)<\frac{1}{2}&\text{ iff }\ \tr\pi\tau<\frac{1}{2}.
\end{align*}

\section{Summary}
\label{sec:discussion}
We have argued that the fidelity of pure states is the quantum
generalization of the classical identity--predicate $\delta_{xy}$,
and showed that an operational basis for it, similar to the classical
way, does not exist. Indeed, there does not exist any quantum
operation behaving like $\delta_{xy}$ on an
orthogonal set of states. Finally, we discussed
the univeral optimal approximation to the fidelity function,
in the simplest case.

\section*{Acknowledgements}
\label{sec:ack}
I would like to thank Cameron Wellard and Serge Massar for conversations
on the present paper's subject, during the QCM\&C 2000 at Capri.

\appendix

\section{The general case}
\label{sec:a:general}
In this appendix we demonstrate a possible attack on the general case.
Unfortunately we find the final optimization problem so hard to solve
that we leave the solution open.
\par
Given $n$ copies of each state we want to produce
as close an approximation to $m$ samples of
$F(\pi,\tau)=(\tr\pi\tau,1-\tr\pi\tau)$ as possible,
i.e. a POVM ${\bf A}$ indexed by $\{0,1\}^m$ which minimizes
$$\delta({\bf A})=\max_{\pi,\tau}
                  \left\|{\bf A}\left(\pi^{\otimes n}\otimes\tau^{\otimes n}\right)
                          -F(\pi,\tau)^{\otimes m}\right\|_1,$$
where we write ${\bf A}\left(\pi^{\otimes n}\otimes\tau^{\otimes n}\right)$
for the distribution on $\{0,1\}^m$ induced by measuring ${\bf A}$
on $\pi^{\otimes n}\otimes\tau^{\otimes n}$.
Obviously we can assume that ${\bf A}$ is supported on
${\cal H}^n_+\otimes{\cal H}^n_+$, where ${\cal H}^n_+$ is the
symmetric subspace in ${\cal H}^{\otimes n}$, i.e. the set of all
vectors invariant under tensor factor permutation.
\par
By the familiar averaging argument we can assume that all elements
of ${\bf A}$ are invariant under the action of the unitary group
${\cal U}({\cal H})$. This action decomposes ${\cal H}^n_+\otimes{\cal H}^n_+$
into $n+1$ orthogonal subspaces ${\cal S}_l$:
the restriction to ${\cal S}_l$ is irreducible
with highest weight $(2n-l,l,0,\ldots,0)$, $l=0,\ldots,n$.
In particular, they all have multiplicity one (for these representation
theoretical details we refer the reader to~\cite{zhelobenko}).
Denote the subspace projection onto ${\cal S}_l$ by $S_l$.
\par
Since $F(\pi,\tau)^{\otimes m}$ has the constant value
$(\tr\pi\tau)^k(1-\tr\pi\tau)^{m-k}$ on the sets
$$\fset{T}_k=\{x^m\in\{0,1\}^m:x^m\text{ has exactly }k\ 0\text{'s}\},$$
we may assume that an optimal ${\bf A}$ is constant on the $\fset{T}_k$
as well. Introducing the angle $\gamma$ between $\ket{\phi}$ and
$\ket{\theta}$, so that $\tr\pi\tau=\cos^2\gamma$ and
$1-\tr\pi\tau=\sin^2\gamma$, we can define
\begin{align*}
  f_k &={\bf A}\left(\pi^{\otimes n}\otimes\tau^{\otimes n}\right)(\fset{T}_k), \\
  p_k &= F(\pi,\tau)^{\otimes m}(\fset{T}_k)
       ={m \choose k}(\cos^2\gamma)^k(\sin^2\gamma)^{m-k},
\end{align*}
and thus write
$$\|{\bf A}\left(\pi^{\otimes n}\otimes\tau^{\otimes n}\right)
                        -F(\pi,\tau)^{\otimes m}\|_1=\sum_{k=0}^m |f_k-p_k|.$$
Observe that with
$$F_k={\bf A}^*(1_{\fset{T}_k})=\sum_{x^m\in\fset{T}_k} A_{x^m}$$
one has $f_k=\tr\left((\pi^{\otimes n}\otimes\tau^{\otimes n})F_k\right)$.
\par
By invariance we can write
$$F_k=\sum_{l=0}^n \alpha_{kl} S_l,\text{ with }
     \alpha_{kl}\geq 0,\quad \sum_{k=0}^m \alpha_{kl}=1.$$
Now, applying invariance once more, we get
$$f_k=\tr\left(\left(\int_{{\cal U}(d)} dU
        (U^{\otimes 2n}\pi^{\otimes n}\otimes\tau^{\otimes n}U^{*\otimes 2n})
         \right)F_k\right).$$
The integral itself is an invariant state, hence of the form
$$\sum_{l=0}^n \beta_l\frac{1}{\tr S_l}S_l,\text{ with }
    \beta_l\geq 0,\quad \sum_{l=0}^n \beta_l=1,$$
and by invariance -- third time pays for all -- the $\beta_l$ depend
solely on $\tr\pi\tau$. In fact, it is easily seen that
they all are homogenous polynomials in $\cos\gamma$ and $\sin\gamma$
of total degree $2n$.
\par
This makes it seem rather unlikely that we can find
$$\delta({\bf A})=\max_{0\leq\gamma\leq\pi/2} \sum_{k=0}^m
               \left|{m \choose k}(\cos^2\gamma)^k(\sin^2\gamma)^{m-k}
                     -\sum_{l=0}^n \alpha_{kl}\beta_l(\cos\gamma,\sin\gamma)\right|,$$
let alone minimize this over the $\alpha_{kl}$.



\begin{thebibliography}{M}
\bibitem{POVM:tomography} Since we want to reconstruct a set of selfadjoint
    operators $(A_1,\ldots,A_n)$ on ${\cal H}$, it is sufficient to know
    the values of $\tr\rho A_i$ for a spanning set of states $\rho$.
    Formally this is the same as state tomography, as it rests on the
    Hermiticity of the Hilbert--Schmidt inner product $\tr A^*B$ for
    operators.
\end{thebibliography}
\end{document}